# Spontaneous Radiation in Relativistic Strophotron

I.V. Dovgan*

Department of Physics, Moscow State Pedagogical University, Moscow 119992, Russia.

**Abstract**. Spectral intensity of spontaneous radiation is calculated in relativistic strophotron. It is shown, that it in the strophotron is given by a superposition of contributions from emission at different (odd) harmonics of the main resonance frequency $\omega_{res}$. The main resonance frequency is shown to depend on the initial conditions of the electron, and in particular on its initial transversal coordinate $x_0$.

## 1. Introduction

In usual Free electron lasers (FEL) electrons move in transverse periodic magnetic field of undulator [1-5]. There are many other systems for creating FEL too. One of them is strophotron [6-11], were electrons oscillate in transverse direction in potential through and travel along strophotron length.

In the present article we investigate spontaneous radiation of electrons in strophotron.

## 2. Spontaneous emission of the strophotron

In the electric strophotron with scalar potential $\Phi(x) = \Phi_0(x/d)^2$, where $\Phi_0$ and $2d$ are the height and the width of the potential "trough" transverse coordinate x has a form [12].

$$x(t) = x_0 \cos \Omega t + \frac{\dot{x}_0}{\Omega} \sin \Omega t \qquad (1)$$

where $x_0$ and $\dot{x}_0 \approx \alpha$ are the initial transversal coordinate and speed of the electron, and frequency of transverse oscillations

$$\Omega^2 = \frac{2e\Phi_0}{\varepsilon_z d^2}. \qquad (2)$$

In accordance with [12], the longitudinal momentum and energy are conserved: $p_z = const$,

*dovganirv@gmail.com

$$\varepsilon_z = \left( p_z^2 + m^2 \right)^{1/2} = const \quad \text{(we use the system of units in which } c = 1).$$

The longitudinal coordinate has an expression [12]

$$z(t) = \dot{z}_0 \left\{ \left( 1 + \frac{\alpha^2 + x_0^2 \Omega^2}{4} \right) t + \frac{\alpha^2 + x_0^2 \Omega^2}{8\Omega} \left[ \sin(2\Omega t + \varphi_0) - \sin \varphi_0 \right] \right\}$$

$$\approx t + \left\{ -\left( \frac{1}{2\gamma^2} + \frac{\alpha^2 + x_0^2 \Omega^2}{4} \right) t + \frac{\alpha^2 + x_0^2 \Omega^2}{8\Omega} \left[ \sin(2\Omega t + \varphi_0) - \sin \varphi_0 \right] \right\} \tag{3}$$

where $\varphi_0 = constant$.

$$\sin \varphi_0 = \frac{2 x_0 \Omega \alpha}{\alpha^2 + x_0^2 \Omega^2}, \qquad \cos \varphi_0 = -\frac{\alpha^2 - x_0^2 \Omega^2}{\alpha^2 + x_0^2 \Omega^2}. \tag{4}$$

Using the solutions (1) and (3), we can find the spectral intensity of a spontaneous emission of the electron. The spectral intensity of emission in the most interesting case, along the $z$ axis, is determined by the well-known formula [13]

$$\frac{d\mathrm{E}_\omega}{d\omega do} = \frac{e^2 \omega^2}{4\pi^2} \left| \int_0^T dt \; \dot{x} e^{i\omega(t-z)} \right|^2 \tag{5}$$

where $do$ is an infinitely small solid angle in the direction $0z$, and T is the time it takes for the electron to travel through the strophotron.

Substituting x(t) (1) and z(t) (3) into (5) and expanding the periodical integrand in the Fourier series, we obtain

$$\frac{d\mathrm{E}_\omega}{d\omega do} = \frac{T^2 e^2 \omega^2}{16\pi^2} \left( \alpha^2 + x_0^2 \Omega^2 \right) \sum_{s=0}^{\infty} \frac{\sin^2 u_s}{u_s^2} \left( J_s(Z) - J_s(Z) \right)^2 \tag{6}$$

where $J_{s,s+1}(Z)$ are the Bessel functions,

$$u_s = \frac{T}{4\gamma^2} \left[ \omega \left( 1 + \frac{\gamma^2}{2} \left( \alpha^2 + x_0^2 \Omega^2 \right) \right) - 2\gamma^2 \Omega(2s+1) \right] \tag{7}$$

$$Z = \frac{\omega}{8\Omega} \left( \alpha^2 + x_0^2 \Omega^2 \right). \tag{8}$$

Equation (6) describes the spectrum of emission consisting of a superposition of the spectral lines located at the odd harmonic $(2s+1)\omega_{res}$ $(s = 0, 1, 2, ...)$ of the main resonance frequency

$$\omega_{res} = \frac{2\gamma^2 \Omega}{1 + \dfrac{\gamma^2}{2} \left( \alpha^2 + x_0^2 \Omega^2 \right)}. \tag{9}$$

These lines are separated each from other by the term $2\omega_{res}$ and their width is found from the condition $u_s = 1$ to be on the order of $2\omega_{res}/\Omega T$. If the number $N_0 = \Omega T = (L/\lambda_0)$ of electron oscillations with strophotron period $\lambda_0 = 1/\Omega$ at the length of the strophotron $L$ is large enough $N_0 \gg 1$, the spectral lines in'(6) for a single electron do not overlap. A relative intensity of emission at the harmonics with a number $(2s + 1)$ is determined by the factor $\left(J_s(Z) - J_s(Z)\right)^2$. If $\alpha\gamma \ll 1$ and $|x_0|\Omega\gamma \ll 1$, $Z = (2s+1)/4\gamma^2\left(\alpha^2 + x_0^2\Omega^2\right) \ll s$. In this case of rather small $\gamma$, $\alpha$, and $|x_0|$, only the main frequency $\omega_{res}$ (9) is emitted with an appreciable intensity. Emission at higher harmonics $s \gg 1$ becomes possible only when $\alpha\gamma \gg 1$ or $|x_0|\Omega\gamma \gg 1$. But, of course, in this case, the main resonance frequency $\omega_{res}$ (9) falls down.

The resonance frequency $\omega_{res}$ (9) has been obtained correctly in the paper [60] (see also [62]).

It is interesting to compare the results (9)-(12) to the corresponding results for the plane undulator [65,68]. These formulas can be reduced to the same form being expressed in terms of the "undulator parameter" $K_{und} = eB_0\lambda_0/2\pi mc^2$ ( $B_0$ and $\lambda_0$. are the magnetic field strength and the period of the undulator) which, in the case of a strophotron, is replaced by

$K_{str} = \alpha\left(\alpha^2 + x_0^2\Omega^2\right)^{1/2} = \lambda(\Omega a/c)$. Now (9)-(12) take the form

$$\frac{d\mathrm{E}_\omega}{d\omega do} = \frac{L^2 r_0 mc^2\gamma^2}{4\pi^2 c^3}\frac{K^2}{\left(1 + \frac{1}{2}K^2\right)}\sum_{s=0}^{\infty}(2s+1)^2\left(J_s - J_{s+1}\right)^2\frac{\sin^2 u_s}{u_s^2} \qquad (10)$$

where $r_0 = e^2/mc^2 2$ is the classical electron radius,

$$u_s = \frac{T}{4\gamma^2}\left[\omega\left(1 + \frac{\gamma^2}{2}\left(\alpha^2 + x_0^2\Omega^2\right)\right) - 2\gamma^2\Omega(2s+1)\right] \qquad (1.4.22\ 11)$$

$$\omega_{res} = \frac{2\gamma^2\Omega}{1 + \frac{1}{2}K^2}. \qquad (1.4.23\ 12)$$

and the argument of the Bessel function Z (8) calculated at $(2s+1)\omega_{res}$ is given by

$$Z = \frac{2s+1}{4}\frac{K^2}{1 + \frac{1}{2}K^2}. \qquad (1.4.24\ 13)$$

Formally, these results completely coincide with the results of the papers [65,68]. Hence, for a single electron, the spectral intensity of a spontaneous emission can be reduced to the coinciding forms in the cases of the strophotron and of the plane undulator. But this coincidence will be shown to disappear when, instead of a single electron, one considers the electron beam as a whole. And the reason is in a different definition and parametrical dependence of the $K$ parameter in the undulator and in the strophotron.

## 3.    Conclusion

Spectral intensity of a spontaneous emission in the strophotron is given by a superposition of contributions from emission or amplification at different (odd) harmonics of the main resonance frequency $\omega_{res}(x_0)$. The main resonance frequency is shown to depend on the initial conditions of the electron, and in particular on its initial transversal coordinate $x_0$. This dependence $\omega_{res}(x_0)$ is shown to give rise to a very strong inhomogeneous broadening of the spectral lines. The broadening can become large enough for the spectral lines to overlap with each other. The obtained results are compared with those for undulator.